\input harvmac

\lref\scqm{J. Michelson and A. Strominger, {\it The Geometry of
(Super) Conformal Quantum Mechanics,} Commun.\ Math.\ Phys.\ {\bf 213} (2000)
1, hep-th/9907191.}

\lref\iceland{R. Britto-Pacumio, J. Michelson, A. Strominger and
A. Volovich, {\it Lectures on Superconformal Quantum Mechanics and
Multi-Black Hole Moduli Spaces,} contributed to NATO Advanced Study
Institute on Quantum Geometry, Akureyri, Iceland,
hep-th/9911066.}

\lref\ddf{V. de Alfaro, S. Fubini and G. Furlan, {\it Conformal
Invariance in Quantum Mechanics,} Il Nuovo Cimento {\bf 34A} (1976)
569.}

\lref\gmt{ J.P. Gauntlett, R.C. Myers and P.K. Townsend,
{\it Black Holes of D=5 Supergravity,} Class.\ Quant.\ Grav.\ {\bf 16} (1999)
1,
hep-th/9810204.}

\lref\gt{G. W. Gibbons and P. K. Townsend, {\it Black Holes and
Calogero Models,} Phys.\ Lett.\ {\bf B454} (1999) 187,
hep-th/9812034.}

\lref\cdkk{P. Claus, M. Derix, R. Kallosh, J. Kumar, P.K. Townsend
and A. Van Proeyen, {\it Black Holes and Superconformal Mechanics,}
Phys.\ Rev.\ Lett. {\bf 81} (1998) 4553,
hep-th/9804177.}

\lref\jmas{J. Michelson and A. Strominger, {\it Superconformal
Multi-Black Hole Quantum Mechanics,} JHEP {\bf 9909} (1999) 005,
hep-th/9908044.}

\lref\gpjg{J. Gutowski and G. Papadopoulos, {\it The Dynamics of Very
Special Black Holes,} Phys.\ Lett.\ {\bf B472} (2000) 45, hep-th/9910022.}

\lref\hull{C. M. Hull, {\it The Geometry of Supersymmetric Quantum
Mechanics,} hep-th/9910028.}

\lref\zanon{S. Cacciatori, D. Klemm and D. Zanon,
{\it $w_{\infty}$ Algebras, Conformal Mechanics, and Black Holes,} Class.\
Quant.\ Grav.\ {\bf 17} (2000) 1731,
hep-th/9910065.}

\lref\bos{T. Banks, M. O'Loughlin and A. Strominger,
{\it Black Hole Remnants and the Information Puzzle,}
Phys.\ Rev.\ {\bf D47} (1993) 4476, hep-th/9211030.}

\lref\pap{G. Papadopoulos, {\it Conformal and Superconformal Mechanics,}
Class.\ Quant.\ Grav.\ {\bf 17} (2000) 3715, hep-th/0002007.}

\lref\mss{A. Maloney, M. Spradlin and A. Strominger,
{\it Superconformal Multi-Black Hole Moduli Spaces in Four Dimensions,}
hep-th/9911001.}

\lref\gutpap{J. Gutowski and G. Papadopoulos,
{\it Moduli Spaces for Four-and Five-Dimensional Black Holes,}
Phys.\ Rev.\ {\bf D62} (2000) 064023, hep-th/0002242.}

\lref\mms{J. Maldacena, J. Michelson and A. Strominger, {\it Anti-de Sitter
Fragmentation,} JHEP {\bf 9902} (1999) 011, hep-th/9812073.}

\lref\kirit{E. Kiritsis, {\it Introduction to Superstring Theory,} Leuven
Univ. Pr. (1998), hep-th/9709062.}

\skip0=\baselineskip
\divide\skip0 by 2

\def\l{\lambda}
\def\tmpsp{\the\skip0}

\let\linesp=\mylinesp
\def\smaleq{&\!\!\!=&\!\!\!}

\def\skipthis#1{{}}

\def\g{\gamma}

\def\O{\Omega}
\def\IR{\relax{\rm I\kern-.18em R}}
\def\IZ{\relax\ifmmode\hbox{Z\kern-.4em Z}\else{Z\kern-.4em Z}\fi}
\def\IQ{\relax{\rm I\kern-.40em Q}}
\def\IS{\relax{\rm I\kern-.18em S}}
\def\p{\partial}

\Title{\vbox{\baselineskip12pt\hbox{hep-th/0004017}\hbox{}
\hbox{}}}{Two-Black-Hole Bound States }

\centerline{
Ruth Britto-Pacumio\foot{{\tt britto@boltzmann.harvard.edu}},
Andrew Strominger\foot{{\tt andy@planck.harvard.edu}}
and Anastasia Volovich\foot{{\tt nastya@physics.harvard.edu}}
\foot{On leave from L. D. Landau Institute for Theoretical Physics,
Moscow, Russia}}
\bigskip\centerline{Department of Physics}
\centerline{Harvard University}
\centerline{Cambridge, MA 02138}

\vskip .3in \centerline{\bf Abstract}
{The quantum mechanics of $N$ slowly-moving BPS black holes in five
dimensions is considered. A divergent continuum of states describing
arbitrarily closely bound black holes with arbitrarily small
excitation energies is found.  A superconformal structure
appears at low energies and can be used to define an index
counting the weighted  number of supersymmetric bound states.
It is shown that the index is determined from the dimensions of
certain cohomology classes on the symmetric product
of $N$ copies of $\IR^4$.
An explicit computation for the case of $N=2$ with no angular momentum yields a
finite nonzero result.
 }
\smallskip
\Date{}
\listtoc
\writetoc

\newsec{Introduction}
The low energy dynamics of $N$ near-coincident
BPS black holes in five dimensions is
described by superconformal quantum mechanics \refs{\jmas \iceland
\gpjg -\gutpap}.
(See also
\refs{\gt\cdkk \mms \zanon \pap-\mss}.)
The
states of this theory describe black holes whose separations and
excitation energies go to zero in the infrared scaling limit.
Hence every state of
this low energy theory describes a marginally bound state of the $N$ black holes.

The existence of infinitely many bound states is puzzling.
In this paper we compute a supersymmetric index which
counts a weighted number of bound states (roughly the difference between the
numbers of hypermultiplets and vector multiplets).
This index,
in contrast, turns
out to be finite and nonvanishing.
After factoring out center-of-mass degrees of freedom,
the index essentially reduces to the Witten index of the black hole quantum mechanics:
\eqn\rfsx{{\cal I}_{BH}(j_L)\equiv Tr(-)^{2J^3_R}|_{j_L}.}
In this expression $J_R$ is the generator of $SU(2)_R$ spatial rotations
and is a generator of the superconformal algebra.  
$SU(2)_L$ spatial rotations commute with the superconformal algebra,
and the trace in \rfsx\ is at fixed total $j_L$.
Ordinarily this would be computed as a trace over
eigenstates of the hamiltonian $H$.
This trace is ill-defined for superconformal quantum mechanics because
of the infrared continuum.  (The usual method of putting the system in a box
does not work here because the continuum arises from near coincident black
holes.)  We define the index by tracing instead
over eigenstates of $L_0=\half(H+K)$, where $K$ is the generator
of special conformal transformations.
For the
case of two black holes we find that ${\cal I}_{BH}(0)=-1$.
 For general $N$ we relate the index to the counting of a certain type of
noncompact cohomology class in the symmetric (for identical black holes)
product of $N$ copies
of $\IR^4$. At higher $N$ there appears to be a
rich structure of supersymmetric bound
states
whose elucidation we defer to later work.

On the way to studying the $N$-black hole problem we describe
some general properties of the Hilbert space of superconformal quantum
mechanics. We begin with the simplest case with two supersymmetries and
work up to black holes.
In section 2 we relate the spectrum of an $Osp(1|2)$ sigma model to
the eigenvalues of a certain Dirac operator on the target space.
In section 3 we show that states of $SU(1,1|1)$
superconformal quantum mechanics are naturally viewed as $(p,0)$-forms,
but with a non-canonical measure, and we derive the  chiral primary condition.
In section 4 we describe the $D(2,1;0)$ quantum mechanics.
In section 5 we explicitly compute the index ${\cal I}_{BH}(0)$ for
two black holes.

\newsec{$Osp(1|2)$}

In this section we relate the spectrum of an $Osp(1|2)$ sigma model to
the eigenvalues of a certain Dirac operator with torsion on the target space.

\subsec{Symmetries}

  The $Osp(1|2)$ superalgebra is generated by the three bosonic operators
$H$, $K$ and $D$ and two fermionic operators $Q$ and $S$. The
nonvanishing commutation relations are
\eqn\ccr{\matrix{\hfill [H,K]\smaleq -iD, \hfill &&&
                 \hfill [H,D]\smaleq -2iH, \hfill &&&
                 \hfill [K,D]\smaleq 2iK,\hfill \linesp
\hfill \{Q,Q\}\smaleq 2H, \hfill &&&
   \hfill [Q,D]\smaleq -iQ, \hfill &&&
   \hfill [Q,K]\smaleq -iS, \hfill \linesp
\hfill \{S,S\}\smaleq 2K, \hfill &&&
   \hfill [S,D]\smaleq iS, \hfill &&&
   \hfill [S,H]\smaleq iQ, \hfill \linesp
\hfill \{S,Q\}\smaleq D.\hfill }}
Alternatively one may define (suppressing an arbitrary  dimensionful
constant)
\eqn\dfg{\eqalign{G_{\pm \half}&= {1 \over \sqrt{2}}(Q\mp iS),\cr
                  L_0&=\half(H+K),\cr L_{\pm 1}&=\half(H-K\mp iD),  }}
in terms of which the algebra becomes
\eqn\algba{[L_m, G_r]={m-2r \over 2} G_{m+r},}
\eqn\algba{\{G_r, G_s\}=2 L_{r+s},}
\eqn\algba{[L_m, L_n]=(m-n)L_{m+n},}
where $r,s=\pm {1 \over 2}$
and $m,n=0,\pm 1.$

The $Osp(1|2)$
algebra can be realized with the supermultiplet $(X^M, \lambda^M)$,
where $\lambda^M=\lambda^{M\dagger}$. The nonvanishing
commutation relations for these
fields and their conjugate momenta are, in the notation of \scqm,
\eqn\erh{\eqalign{\{\lambda^M,\lambda^N\}&=g^{MN},\cr
                 [P_M,X^N]&=-i\delta_M^{~~N},\cr
                  [P_M,\lambda^N]&=i(\Gamma_{~~MP}^{N}-\omega_{M~~P}^{~~~N})
                \lambda^P,}}
with $\omega$ the spin connection and $\Gamma$ the Christoffel connection.
The last relation is necessitated by the fact that the commutator of
$\lambda$ with itself depends on $X$, and implies that
$e^\alpha_{~M}\lambda^M$,
where $e$ is the vielbein, commutes with $P$.
Explicit expressions for the supercharges
are then
\eqn\sscq{Q=\lambda^MP_M +{i \over 6}(c_{MNP}-3\omega_{MNP}) \lambda^M
                  \lambda^N \lambda^P,}
and
\eqn\sdj{S=\lambda^MD_M.}
The algebra requires that the vector field $D$ is a so-called closed
homothety obeying
\eqn\ldrf{\eqalign{ {\cal L}_D g_{MN}&=2g_{MN},
\cr                       d (D_MdX^M)&=0, }}
and the torsion $c$ obeys
\eqn\rtok{\eqalign{D^Mc_{MNP}&=0,\cr {\cal L}_Dc_{MNP}&=2c_{MNP}.} }
A derivation of these results can be found in \scqm.

\subsec{Quantum States}
States $|\psi\rangle$ in the Hilbert space form
a representation of the algebra of
fermions. From the commutation relations \erh, we conclude that the states
are target space spinors, with the fermions acting as
\eqn\rtp{\lambda^M |\psi\rangle ={1 \over \sqrt{2}}\gamma^M|\psi\rangle,}
where $\gamma^M$ are the usual $SO(N)$ gamma matrices.

The states can be organized into infinite-dimensional
superconformal multiplets. At the bottom
of every multiplet is a superprimary state obeying
\eqn\spr{G_\half |\psi\rangle =0.}
The remaining tower of states is generated by the action of
$G_{-\half}$.
Superprimary states are related by a similarity transformation to
states $|\psi^\prime\rangle =e^{K}|\psi\rangle$ obeying
\eqn\ftl{Q|\psi^\prime\rangle=0.}
Using \rtp, \ftl\ becomes the modified Dirac equation
\eqn\hik{-i(\gamma^M\nabla_M-{1 \over 12}\gamma^{~~NP}_Mc^M_{~~NP})
|\psi^\prime\rangle=0.}
In order to better understand \hik, we introduce the
conformally related metric
\eqn\str{d\tilde s^2= K^{-1}ds^2.}
It follows from \ldrf\ that the vector field $D$ is covariantly constant in
this metric. Hence coordinates can be chosen so that it takes the simple
product form
\eqn\sfl{d\tilde s^2=2(dX^0)^2+\hat g_{IJ}dX^IdX^J,}
where the $(N-1)$-dimensional metric $\hat g$ on the space
$M^{N-1}$ transverse to the orbits of $D$ is independent of
$X^0$. In these coordinates, $D^M\p_M=\p_0$ and $K=\half e^{2X^0}$.
Equation \ftl\ then reduces to
\eqn\higk{-i(\gamma^I\hat\nabla_I-{1 \over 12}\gamma^{~~JK}_Ic_{~~JK}^I)
|\psi^\prime\rangle=
i\gamma^0(\p_0+{N-1 \over 2})|\psi^\prime\rangle. }  The left hand side of \higk\ is
a Dirac equation with torsion on
the transverse space $M^{N-1}$. Let $\lambda_{j}$ denote the (real)
spectrum of this
Dirac operator and $\hat \psi_{j}$ the corresponding orthonormal eigenspinors.
In an even-dimensional space, there is a basis in which
\eqn\gam{
\gamma^{I}=
\pmatrix{0 & \gamma^{I} \cr \gamma^{I} & 0},~~~
\gamma^{0}=
\pmatrix{0 & -i {\bf 1}  \cr i {\bf 1} & 0},}
and a solution of the Dirac equation can be written as
\eqn\resh{
|\psi'_i \rangle=
{\psi_U' \choose \psi_L'}=
{
\hat\psi_i e^{-({N-1 \over 2}+ \lambda_i)X^0} \choose
\hat\psi_i e^{-({N-1 \over 2}- \lambda_i)X^0}
}.
}
Superprimaries obeying \spr\ are obtained by a similarity transformation as
\eqn\solns{
|\psi_i\rangle=
{\psi_U \choose \psi_L}=
{\hat\psi_i e^{-{1 \over 2}e^{2X^0}  -({N-1 \over 2}+ \lambda_i)X^0} \choose
\hat\psi_i e^{-{1 \over 2}e^{2X^0} -({N-1 \over 2}- \lambda_i)X^0}}.
}
Normalizability then requires convergence of the integral
\eqn\conv{\langle\psi_j|\psi_k\rangle=\int d^NX\sqrt{g} \psi^\dagger_j\psi_k
=\delta_{jk}2^{{1-N} \over 2}\int dX^0 (2\sinh 2 \lambda_j X^0)
e^{[X^0-e^{2X^0}]}, }
which is equivalent to the condition that $|\lambda_j|<\half.$
We conclude that there is one superprimary for every such
normalizable solution of the Dirac equation with torsion
on the transverse space $M^{N-1}$.


\newsec{$SU(1,1|1)$}

In this section we show that states of $SU(1,1|1)$
superconformal quantum mechanics are naturally viewed as $(p,0)$-forms,
but with a non-canonical measure. The chiral primary condition
is derived and expressed as a condition on forms.

\subsec{Symmetries}

The $SU(1,1|1)$ algebra is
\eqn\rtal{[L_m,L_n]=(m-n)L_{m+n}}
\eqn\rtbl{\{G_r,\bar{G}_s\}=2L_{r+s}+2 J(r-s)\delta_{r,-s}}
\eqn\rtcl{[J, G_r]={1 \over 2}G_r, ~~~[J, \bar{G}_r]=-{1 \over 2}\bar{G}_r}
\eqn\rtdl{[L_m,G_r]={m-2r \over 2}G_{m+r}, ~~~
[L_m,\bar{G}_r]={m-2r \over 2}\bar{G}_{m+r}}
where $m,n=0,\pm 1$ and $r,s=\pm\half$.
As shown in \scqm, the $Osp(1|2)$ model of
the previous section has this larger
symmetry if and only if there is a complex structure preserved by
the action of $D$,
the metric is hermitian, and the $(1,2)$ part of the torsion is
given by
\eqn\cgm{c^{\bar a}_{~~\bar b c}=-
\Gamma^{\bar a}_{~~\bar b c}. }
Indices in lower (upper) case denote complex (real) coordinates,
so that $a,b=1,2,\ldots,n$, where $n$ is the complex dimension.
In general $c$ may also have
$(0,3)$ and $(3,0)$ parts unconstrained by \cgm.
These are constrained to vanish for
${\cal N}=4$ supersymmetry, and for simplicity we set them to zero here.
The relation
\cgm\ then implies that there is a $U(n)$ connection,
\eqn\unc{ \Omega^N_{~MP} \equiv \Gamma^N_{~MP}+c^N_{~MP}.}

Let us first collect some formulae from \scqm\ describing the $SU(1,1|1)$
theories.\foot{Our notation is the same as in \scqm\ with the following
exceptions.  We use capital indices $M,N,P$ for the
$2n$-dimensional moduli space coordinates.  The indices $A,B$ are used only for the black
holes themselves.  In this section,
$Q$ and $S$ are the holomorphic supercharges that were called ${\cal
{Q}}={1 \over 2}(Q -i {\tilde Q})$
and ${\cal {S}}={1 \over 2}(S-i\tilde{S})$ in \scqm.  In section 4, $R$ and $J_R$ are the
angular momentum operators corresponding to $R_+$ and $R_-$ in \scqm.}
It is convenient to define the shifted momentum
\eqn\pdf{\Pi_M=P_M-{i \over 2}(\omega_{MNP}-c_{MNP})\l^N\l^P ,}
which has the property
\eqn\pipo{[\Pi_M,\l^N]=i\Omega^N_{~MP}\l^P.}
Defining $Q$, $S$ and their hermitian conjugates $\bar Q$, $\bar S$ by
\eqn\tfy{\eqalign{G_{\pm \half}&={1 \over \sqrt{2}}(Q\mp iS),\cr
\bar G_{\pm \half}&={1 \over \sqrt{2}}(\bar Q\mp i\bar S),\cr}}
and
\eqn\rto{c_a={c^b}_{ba},}
one has
\eqn\qdf{\eqalign{Q&=\l^a(\Pi_a-ic_{ab\bar c}\l^b\l^{\bar c}-ic_a),  \cr
S&=\l^aD_a. }}
We note that
\eqn\eri{\eqalign{\{ Q,\l^a\}&=0,  \cr
   \{ Q,\l^{\bar a} \}&=g^{\bar a b}(\Pi_b-ic_b). }}

\subsec{The Ground State of $H$}

In this subsection we construct a state $|\eta\rangle$ annihilated
by the supercharges $Q$ and $\bar Q$ as well as the Hamiltonian
$H$.  We compute the norm of $|\eta\rangle$ and its $U(1)$ charge.

We begin by imposing the chirality conditions \eqn\ccd{\g^{\bar
a}|\eta\rangle=\g_a|\eta\rangle=0,} which states that
$|\eta\rangle$ is a singlet under the $SU(n)$ subgroup of $U(n)$.
It also implies that $|\eta\rangle$ is annihilated by $\bar Q$ (as
well as $\bar S$). The action of the supercharge $Q$ on a general
state $|\psi\rangle$ is \eqn\rtol{Q|\psi\rangle=-i\g^a({\cal
D}_a+c_{ab\bar c}\g^b\g^{\bar c}+c_a)|\psi\rangle,} where $\cal D$
is the covariant derivative on spinors with the $U(n)$ spin
connection associated to $\O$. Given \ccd, the state
$|\eta\rangle$ will be annihilated by $Q$ if the wavefunction
$\eta(X)$ is a covariantly constant spinor obeying
\eqn\erty{({\cal D}_a+c_a)\eta =0.} This has solutions because the
integrability conditions $[ {\cal D}_a+c_a,{\cal D}_b+c_b]\eta
=-2c^c_{~~ab}({\cal D}_c+c_c)\eta$ are satisfied, as can be checked
from the expression \unc\ for  the connection. The solution is
fixed up to transformations of the form \eqn\ser{ \eta \to
f(X^{\bar a})\eta,} where $f$ is an antiholomorphic function. We
will fix this freedom shortly.

We now describe some properties of $|\eta\rangle$.
Define the $U(1)$ connection $A$ by
\eqn\ucn{A_N=\Omega^M_{~NP}I_M^{~~P}=i\Omega^a_{~Na}-i\Omega^{\bar a}_{~N\bar a} ,}
where $I$ is the complex structure.
$A$ obeys
\eqn\rtl{A_b=-4ic_b+2i\p_b \phi, }
with
\eqn\gtm{\phi={1 \over 4} \ln \det g .}
Under an infinitesimal complex coordinate transformation
\eqn\rpm{X^a \to X^a+\zeta^a(X^b),}
one finds
\eqn\tkn{\eqalign{\phi &\to \phi + { i \over 2}(\epsilon -\bar \epsilon) \cr
 A &\to A-d (\epsilon+\bar\epsilon), }}
with complex holomorphic gauge parameter
\eqn\epss{\epsilon = i \p_a\zeta^a. }

Using the fact that the spinorial wavefunction  $\eta$ is an
$SU(n)$ (but not a $U(1)$) singlet, the equation \erty\ can be
written, using \rtl , as \eqn\rtkk{(\p_a-{i \over 4}A_a+c_a)\eta
=(\p_a+{1 \over 2}\p_a \phi)\eta=0.} The solution of this equation is
\eqn\sltf{\eta=e^{-{\phi \over 2}+\bar \Theta}\eta_0,} where $\bar
\Theta$ is an arbitrary antiholomorphic function, and $\eta_0$ is
the constant spinor obeying \ccd. The norm of $\eta$ is then
\eqn\trp{\eta^\dagger\eta=e^{-{\phi}+\Theta +\bar \Theta}. }

The expression \trp\ is coordinate invariant because $\Theta +\bar
\Theta$ and $\phi$ transform the same way under \rpm.  In the
following it will be convenient to choose coordinates so that
$\phi$ is nonsingular at smooth points in the geometry. (In the
${\cal N}=4$ case, such coordinates are singled out by the
existence of a quaternionic structure.) The norm \trp\ is then in
general singular, except if the freedom \ser\ is used to shift
away $\Theta$ altogether. We shall henceforth assume that this has
been done, so that in nonsingular coordinates
\eqn\trp{\eta^\dagger\eta=e^{-{\phi} }. }

The $U(1)$ $R$-charge is measured by the operator \eqn\rch{J={1 \over
2} \bigl(iD^a\Pi_a-iD^{\bar a}\Pi_{\bar a}+g_{a \bar b}(\l^a
\l^{\bar b} -\l^{\bar b}\l^a)+D^ac_{aMN}\l^M\l^N-D^{\bar a}c_{\bar
aMN}\l^M\l^N\bigr) .} Acting on $|\eta\rangle$, one finds
\eqn\rtg{J|\eta\rangle= {1 \over 2}(D^{\bar a} \p_{\bar a}{\phi}
-n) |\eta\rangle. } In dilational gauge, defined by
\eqn\dgg{D^M={2 \over h}X^M,} for some constant $h$, one has
\eqn\bgo{ D^{\bar a}\p_{\bar a}\phi={n(h-2) \over 2 h},} and
\eqn\gtol{J|\eta\rangle=-n{h+2 \over 4h} |\eta\rangle , } where
$n$ is the complex dimension of the target space.

\subsec{Chiral Primaries and $(p,0)$-Forms}

In the previous subsection we constructed a supersymmetric ground
state $|\eta\rangle$ of the Hamiltonian $H$. In general this state
may not be normalizable due to the noncompact regions of the
target space.  $L_0$ eigenstates will in some cases be
normalizable. In this subsection we build such states by acting on
$|\eta\rangle$ with bosonic and fermionic operators.

As in the case of $Osp(1|2)$,  $L_0$ eigenstates lie in
superconformal representations containing lowest weight states
annihilated by $G_{\half}$ and $\bar G_{\half}$.  For $SU(1,1|1)$
there are special representations whose lowest weight states are
also annihilated by $G_{-\half}$ (or $\bar G_{-\half}$). These are
chiral (or antichiral) primary states.

Consider a similarity transformation from $|\eta\rangle$ to the
state \eqn\puy{|0\rangle= e^{-K}|\eta\rangle .} Using \eqn\cdf{
G_{\half}={1\over\sqrt{2}}e^{-K}Qe^K,} and \ccd, this state is seen to
obey \eqn\drf{G_{\half}|0\rangle= \bar G_{\pm \half}|0\rangle=0.}
Hence it is a lowest weight antichiral primary. We shall see later
that this state is not normalizable in the black hole case.

The most general state is of the form \eqn\gst{|f_p\rangle
=f_p|0\rangle,} with \eqn\sdfw{f_p={1 \over p !}f_{a_1
a_2...a_p}\l^{a_1}\l^{a_2}...\l^{a_p},} where $f_{a_1 a_2...a_p}$
is totally antisymmetric. Hence the Hilbert space can be
identified with the space of $(p,0)$-forms on the target space.
The action of the supercharges on these states is\foot{The use of a capital
index in the second equation produces one extra term.  For example,
$\nabla^Mf_M=g^{a\bar b}(\p_{\bar b}f_a-2\Gamma^c_{~\bar b a}f_c)$, while
$\nabla^af_a=g^{a\bar b}(\p_{\bar b}f_a-\Gamma^c_{~\bar b a}f_c)$.} 
\eqn\tyo{(Q-iS)f_p|0\rangle =-{i \over p !}(\p_{a_1}f_{ a_2...a_{p+1}})
\l^{a_1}\l^{a_2}...\l^{a_{p+1}}|0\rangle, } 
\eqn\tyeo{(\bar Q-i\bar S)f_p|0\rangle=-{i \over (p-1) !}(e^{{\phi}}\nabla^{M}e^{-{\phi}}f_{Ma_2...a_p})\l^{a_2}...\l^{a_p}|0\rangle,}
\eqn\tdo{Sf_p|0\rangle ={1 \over p !}
D_{a_1}f_{a_2...a_{p+1}}\l^{a_1}...\l^{a_{p+1}}|0\rangle, }
\eqn\tyso{\bar Sf_p|0\rangle={1 \over (p-1) !}D^{a_1}f_{a_1
a_2...a_p}\l^{a_2}...\l^{a_p}|0\rangle. } Regarding $f_p$ as a
$(p,0)$-form, such states are chiral primary if and only if
\eqn\chprc{\p f_p= D \wedge f_p=\bar \p *e^{-{\phi}}f_p=0, } where
$*$ is the Hodge dual and here $D=D_adX^a$ is a $(1,0)$-form.

Note that the inner product,  \eqn\tfl{\langle
f^{\prime}_p|f_p\rangle={1 \over p !}\int d^{2n}x\sqrt{g} e^{-\phi-2K}\bar
f^{\prime
 a_1  a_2 ...  a_p}f_{
 a_1  a_2 ...  a_p}, }
contains extra factors in the integration measure.

\subsec{$Tr(-)^{2J}$}

In this subsection we consider the Witten index \eqn\ftyc{\chi =
Tr(-)^{2J}.} In order to compute the
trace one must choose a basis for the Hilbert space and a
regulator for the infinite sum. One might attempt to define the
trace as the limit of a weighted sum over $H$ eigenstates
\eqn\trdf{\chi=Tr_H(-)^{2J}e^{-\beta H},  ~~~~\beta \to 0.} Of
course the sum is actually independent of $\beta$, since states
with nonzero $H$ come in bose-fermi pairs which cancel. The trace \trdf\ is
nevertheless difficult to evaluate because of the continuum of
eigenstates extending down to zero energy. An alternate way to
define the index is as a weighted sum over $L_0$
eigenstates\foot{In 1+1 dimensions, (3.45) is exactly the
expression obtained for the Witten index in the NS sector obtained
by spectral flow from the R sector. In 0+1 there is no obvious
analog of spectral flow.}
\eqn\tdf{\chi=Tr_{L_0}(-)^{2J}e^{-\beta(L_0-J) },  ~~~~\beta \to
0.} The sum is also independent of $\beta$. This follows from
considering the operator \eqn\opq{\tilde Q= {1 \over \sqrt{2}}(\bar
G_{\half}+G_{-\half}) } with the properties \eqn\topl{\tilde Q^2
=L_0-J,  ~~~~~~~~~\{ \tilde Q, (-)^{2J} \}=0. } States with nonzero
$L_0-J   $ come in bose-fermi pairs generated by $\tilde Q$
and cancel in the sum \tdf.
Hence $\chi$ receives contributions only from states with $L_0-J=0
$. Such states are chiral primaries annihilated by $\tilde Q$, or
equivalently, states annihilated by $\bar G_{\half}$ and $G_{\pm\half}$. Hence the Witten
index can be computed as a weighted sum over superconformal
chiral primaries.

A more general index can be defined when the theory contains an operator
$\cal O$, usually associated with a symmetry, which commutes with the
generators of the superalgebra. In that case the preceding 
argment may be repeated to show that 
\eqn\tdsf{\chi_{\cal O}
=Tr_{L_0}(-)^{{2J}}{\cal O}e^{-\beta(L_0-J) },  ~~~~\beta \to
0}
also defines an index. Alternatively 
the index can be restricted to a sum over the eigenstates of 
$\cal O$ with eigenvalue  $\lambda$
\eqn\tdf{\chi(\lambda)
=Tr_{L_0}(-)^{{2J}}e^{-\beta(L_0-J) }|_\lambda,  ~~~~\beta \to
0}
\newsec{$D(2,1;0)$ }

   In this section we describe $D(2,1;0)$ quantum mechanics in terms of its
symmetries and chiral primaries.  This algebra contains an $SU(1,1|1)$
subalgebra and will be used in section 5 to describe the theory of $N$ BPS black holes.

\subsec{Symmetries}

 $D(2,1;0)$ is the semidirect product of
$SU(1,1|2)$ and $SU(2)_R$. In $4k$ dimensions the target space
geometry has a triplet of self-dual complex structures obeying
\eqn\rtyo{I^rI^s=-\delta^{rs}+\epsilon^{rst}I^t,} for $r,s=1,2,3$.
There are also isometries generated by
\eqn\rtgi{D^{rM}=D^NI^{rM}_N,} whose associated charges generate
the $R$-symmetry  in $SU(1,1|2)$.  This $R$-symmetry will be denoted
by $SU(2)_{\rm Right}$. A large class of $D(2,1;0)$ theories are
sigma models with target space metrics \eqn\met{g_{a \bar
b}=\half\bigl(\p_a\p_{\bar b}L + I^{-\bar c}_aI^{+d}_{\bar
b}\p_{\bar c}\p_d L).} In this expression, $L$ is a homogeneous
function of degree $-2$, the complex coordinates are adapted to
$I^3$, and $I^\pm=\half(I^1\pm i I^2)$.  All three complex
structures are constant self-dual matrices, and $D^M=-X^M$.

It follows from \met\ that \eqn\ftl{\p_a\phi=2c_a.} Hence for
$D(2,1;0)$ the
$U(1)$ connection $A$ in \ucn\ vanishes, and $\Omega$ is an
$SU(n)$ (rather than $U(n)$) connection. The holomorphic two-form $I^-$ obeys
the relations \eqn\daz{\p I^-=0, ~~~~\bar \p * e^{-\phi} I^-=0.}

There are 8 supercharges in $SU(1,1|2)$,
\eqn\rty{G^{\pm \pm}_{\pm \half}={1 \over \sqrt{2}}
(Q^{\pm \pm} \mp iS^{\pm \pm}) .}
Explicit expressions for four of these are
\eqn\gexp{\eqalign{G^{+-}_{\pm \half}&=\l^a(\Pi_a-ic_{ab\bar c}\l^b\l^{\bar
c}-ic_a \mp i D_a),  \cr
G^{-+}_{\pm \half}&=\l^{\bar a}(\Pi_{\bar a}-ic_{{\bar a}b\bar c}\l^b\l^{\bar
c}-ic_{\bar a} \mp i D_{\bar a}). }}
These charges are $SU(2)_{\rm Right}\times SU(2)_R$ doublets.
The $SU(2)_{\rm Right}$
($SU(2)_R$) spin is
indicated by the first (second) superscript.
The other four supercharges can accordingly be expressed as
\eqn\phase{G^{++}_{\pm \half} = i [J^+_R,G^{-+}_{\pm \half}] = -i [R^+,G^{+-}_{\pm \half}],}
\eqn\cphase{G^{--}_{\pm \half} = i [J^-_R,G^{+-}_{\pm \half}] = -i [R^-,G^{-+}_{\pm
\half}].
}

Some important commutators are

\eqn\gcr{\eqalign{\{G^{\alpha \alpha^\prime}_p, G^{\beta\beta^\prime}_q \}
&=2\delta^{\alpha,- \beta  }\delta^{\alpha^\prime,- \beta^\prime  }L_{p+q}
+2(p-q)J_R^{\alpha \beta  }\delta^{\alpha^\prime, -\beta^\prime},  \cr
[J^3_R,G^{\alpha \alpha^\prime}_p]&={\alpha \over 2}
G^{\alpha \alpha^\prime}_p,  \cr[R^3,G^{\alpha
\alpha^\prime}_p]&={\alpha^\prime
   \over 2}  G^{\alpha \alpha^\prime}_p,  }}
with $p,q=\pm\half$, $\alpha \beta  =\pm$ and
$J^{+-}_R=-J^{-+}_R=J^3_R$ etc. $J_R$ generates $SU(2)_{\rm
Right}$ rotations of the bosons $X^A$,
\eqn\frg{[J_R^r,\lambda^M]=0, ~~~~~~~~~~~~~~~~ [J_R^r,X^M]={i
\over 2} X^NI^{rM}_N.} The second $SU(2)_R$, which does not lie in
$SU(1,1|2)$, generates $R$-symmetry transformations of the
fermions.\foot{ $(R,J_R)$ are the operators $(R_+, R_-)$ of \scqm.}
\eqn\ods{ [R^r,\lambda^M]={i \over 2}
\lambda^NI^{rM}_N,~~~~~~~~~~~~~~~~[R^r,X^M]=0.} The supercharges
transform in the $(2,2)$ of $SU(2)_{\rm Right}\times SU(2)_R$.

\subsec{Chiral Primaries }

Chiral primaries in these theories are defined by embedding an
$SU(1,1|1)$ subalgebra in $D(2,1;0)$. A natural embedding is defined by
\eqn\try{G_{\pm \half}=G^{+-}_{\pm\half},~~~~\bar
G_{\pm \half}=G^{-+}_{\pm\half},~~~~J=J^3_R.}
With this embedding, the conditions for a chiral primary are given by
\chprc, in the complex structure defined by $I^3$.

It follows from the algebra that normalizable chiral primaries
annihilated by $\bar G_\half$ and $G_{\pm \half}$ as in \drf\ are
annihilated by six of the eight supercharges
\eqn\rglss{G^{++}_{\pm\half},~~~~G^{+-}_{\pm\half},~~~~G^{-+}_{\half},
~~~~G^{--}_{\half}.} Antichiral primaries are annihilated by the
complementary set of negatively moded supercharges.  Note that
$SU(2)_R$ does not mix the operators in \rglss\ with
$G^{-\pm}_{-\half}$. Therefore if a chiral primary is not an
$SU(2)_R$ singlet, then the $SU(2)_R$ action gives new chiral primaries.
The algebra further implies that all chiral primaries are highest
weight $SU(2)_{\rm Right}$ states annihilated by
$J^+_R$.

We note that the state $|0 \rangle$ is annihilated by all the
positively moded supercharges as well as $G^{\pm +}_{-\half}$.
Hence it is neither a chiral nor an antichiral primary. This implies
it cannot be a normalizable state, as indeed can be seen from
explicit computation. Since $h=-2$ for $D(2,1;0)$, it follows from
\gtol\ that $J^3_R|0\rangle=0$, while $R^3|0\rangle={n \over 4 }|0\rangle$.

\newsec{Black Hole Quantum Mechanics}
At low energies the quantum mechanics of $N$ black holes is
described by the product of a free theory containing the
center-of-mass coordinates and an interacting near-horizon
superconformal theory with a $4(N-1)$-dimensional target space. In
this section we describe this theory and the index ${\cal I}_{BH}$
that counts the weighted number of supersymmetric bound states.
For the case of $N=2$ we find all the $j_L=0$ chiral primaries and
discover ${\cal I}_{BH}(0)=-1$.

\subsec{The Center-of-Mass Multiplet}

Throughout this paper we have largely neglected the
 center-of-mass degrees of freedom.  We give a brief 
description here for completeness.
The center-of-mass theory contains two complex bosons $X^k$ in the
(2,2) of $SU(2)_{\rm Left} \times SU(2)_{\rm Right}$ and two
complex fermions $\l^k$ in the (2,2) of $SU(2)_{\rm Left} \times
SU(2)_R$. The ground state is therefore a spinor of $SU(2)_{\rm
Left} \times SU(2)_R$. Explicitly, defining $|\eta\rangle$ as a zero
momentum state obeying \eqn\gai{\l^{\bar k}|\eta\rangle=0,} there is an
$SU(2)_R$ doublet of spacetime bosons \eqn\exn{|\eta\rangle,~~~~~
\l^1\l^2|\eta\rangle,} and an  $SU(2)_{\rm Left}$ doublet of spacetime
fermions \eqn\exn{\l^1|\eta\rangle,~~~~~ \l^2|\eta\rangle.} This is exactly
the content of a massive, positively charged, spacetime
hypermultiplet. These states are all annihilated by the four
supercharges $Q^\alpha$ because $P_k=0$.

\subsec{The Superconformal Sector}

The low energy interactions between $N$ five-dimensional BPS black
holes with charges $Q_A$ are described by the $4(N-1)$-dimensional
$D(2,1;0)$ theory with metric \met\ constructed from the potential
\jmas \eqn\gtf{L=-\int d^4 X \bigl( \sum_{A=1}^N {Q_A \over |\vec
X- \vec X^A|^2}\bigr)^3.} In this context $SU(2)_{\rm Right}$ in
$SU(1,1|2)$ generates right-handed spatial rotations \gmt, while
$SU(2)_R$ generates the spacetime $R$-symmetry
transformations.

The theory following from \gtf\  has additional global symmetries
corresponding  to $SU(2)_{\rm Left}$ spatial rotations. These are
generated by \eqn\gtn{J^r_L=-{1 \over 2}X^M K^{r~N}_M(\Pi_N - { i
\over 2}\Omega^P_{~NQ}\l^Q\l_P)+{i \over 4}K^r_{MN} \l^M\l^N,}
where $K^r$ are the constant anti-self dual complex structures on
$\IR^4$. $J_L$ obeys \eqn\jki{[J^r_L,\lambda^M]={i \over 2}
\lambda^NK^{rM}_N,~~~~~~~~~~~~ [J^r_L,X^M]={i \over 2}
X^NK^{rM}_N~~~~~~~~~~~~[J^r_L,J^s_R]=0,} and commutes with all the
generators of the superconformal group. $\l^M$ transforms in the
$(1,2,2)$ of $SU(2)_{\rm Right}\times SU(2)_R\times SU(2)_{\rm
Left}$, as appropriate for a goldstino. The full symmetry group of
the system is $SU(2)_{\rm Left}\times D(2,1;0)$.

\subsec{Validity of the Approximations}

In this subsection we discuss potential corrections to
the theory defined by \gtf, and in particular whether or not they
could affect the conclusion that there is a divergent continuum
of infrared states.

The superconformal theory defined by \gtf\ was derived in \jmas\
from  a more general (non-superconformal)
quantum mechanics, in which a constant is added to the sum
inside the parentheses. \gtf\ then arises in an
$M_p \to \infty$ limit \refs{\mms,\jmas}, with the rescaled separations
$|\vec X^A-\vec X^B|\to M_p^{3/2}|\vec X^A-\vec X^B|$
(with dimensions $\sqrt{mass}$) held fixed. At the same time the energies
are rescaled by a factor of $M_p$ so that, in the limit, all
excitations of the superconformal theory have zero energy as measured
with respect to the original time coordinate at
spatial infinity. We wish to know whether all of these
zero-energy states are really
present or whether some of them might be removed by corrections which have been
neglected so far.

  Since $M_p$ has been taken to infinity, there can be no $1/M_p$
corrections to \gtf. As there are no
dimensionful parameters in the infrared limit, corrections to \gtf\
must be supressed by dimensionless quantities such as
$\dot X/X^3$. Such terms can indeed be seen to arise
for example as Born-Infeld type corrections. For the case of two black
holes the corrected action is of the general form
\eqn\fto{\half \int dt\bigl( {|\p_t\vec X|^2 \over |\vec X|^4}+{|\p_t \vec
X|^4 \over |
\vec X|^{10}}....\bigr),}
where $\vec X$ is the relative separation.
In terms of the momentum defined from the leading term
\eqn\oda{\vec P={\p_t \vec X \over |\vec X|^4},}
this becomes
\eqn\fto{\half \int dt\bigl( { |\vec X|^4 |\vec P|^2}+{|\vec X|^6|\vec P|^4}+....\bigr).}
Such correction terms can be neglected as long as
\eqn\ftlac{|\vec P|\ll{1 \over |\vec X|}.}
As the black holes approach one another, the relative momenta must be
smaller and smaller in order to suppress corrections.
Hence we do not expect all states to be reliably described
by the superconformal theory.

The number of states which can be reliably described
by the superconformal theory can be estimated by the volume
$\Omega$ of phase
space in which \ftlac\ is obeyed.
This is, with an infrared cutoff
 $\epsilon \to 0$,
\eqn\mgsa{\Omega\sim\int_{|\vec X|>\epsilon} d^4X \int_{|\vec P|<1/|\vec X|}d^4P \sim
\ln \epsilon.}
Hence, according to this rather crude estimate,
 a logarithmic infrared
divergence in the number of states appears to remain even when
the untrustworthy regions of phase space are removed.

We can also compute the density of states as function of the
energy. This is
\eqn\rkaz{d\Omega \sim \int d^4X d^4P \delta (E-|\vec X|^4|\vec P|^2)dE.}
This is divergent for any $E$
if one does not impose the restriction
\ftlac.  Imposing \ftlac\ leads to the finite, scale invariant, result,
\eqn\fzsd{d\Omega \sim {dE \over E}.}
Since this density of (reliably present)
states is finite, it is possible that the
infrared divergences do not appear in physical
processes involving scattering off of the collection of
black holes.
A similar mechanism was discussed in \bos.

Another potential source of corrections comes from
black hole fragmentation as in \mms. The moduli space geometry
\met , \gtf\ was derived using the low-energy
supergravity approximation. The validity of this requires
that the spacetime curvature is small compared to the
inverse Planck length,
or equivalently $Q_A \gg 1$.\foot{Perhaps surprisingly,
if $Q_A\gg 1$, then the curvatures remain small even
for $\vec X^A-\vec X^B \to 0$.} In this paper we have ignored the
possibility that the black holes might fragment into smaller
pieces. In five dimensions we know of no way to suppress this
energetically. One might try to avoid this by taking all the black
holes to carry the minimum quantum of charge. However it is not clear whether
the expression for the moduli space metric remains valid for
small charges. The expression is highly constrained both by the symmetries
and the known long-distance behavior. Whether or not corrections
do appear at small charge is an open question which we shall not
attempt to resolve here.

In four dimensions the situation is better. Let all the
black holes carry the same charges, with large, nonzero
coprime electric and magnetic charges.  In that case
the possiblity of fragmentation is eliminated energetically,
as can be seen from the BPS mass formula. The moduli
space geometry \refs{\mss,\gutpap} and the bound-state
analysis are similar for this case.

\subsec{${\cal I}_{BH}$}

In this section we relate a spacetime index counting
weighted degeneracies of spacetime BPS multiplets
to an index in the superconformal quantum mechanics
of the type discussed in subsection 3.4.

Massive representations of ${\cal N}=2$ Poincar\'e supersymmetry in
five dimensions are of two types. The generic representation
is the long multiplet $L_j$, with $SU(2)_{\rm Left}\times
SU(2)_{\rm Right}$ spin content
\eqn\ljk{L_j :~~~[j_L,j_R]\otimes([1/2,1/2]+2[1/2,0]+2[0,1/2]+4[0,0]).}
This multiplet has $8(2j_L+1)(2j_R+1)$ bosons,  $8(2j_L+1)(2j_R+1)$ fermions,
and maximal spin $(j_L+1/2,j_R+1/2) $.  There is also a short
multiplet $S_j$ which is annihilated by half of the supercharges
and has $Mass=Charge$ in appropriate units.
This multiplet has spin content
\eqn\sljk{S_j :~~~[j_L,j_R]\otimes([1/2,0]+2[0,0]).}
This multiplet has one quarter\foot{We do not include the conjugate
multiplet with negative charges.} as many bosons and fermions,
and maximal spin $(j_L+1/2,j_R)$. It is easy to see that for either
multiplet
\eqn\erfx{Tr(-)^{2J^3_L+2J^3_R}=0.}
An alternative index,
\eqn\erfsx{{\cal I}\equiv Tr(-)^{2J^3_R}y^{2J^3_L},}
vanishes for long multiplets
\eqn\dzx{{\cal I}(L_j)=0,}
but not for short ones:
\eqn\dazx{{\cal I}(S_j)=(-)^{2j_R}(2j_R+1){(y^\half+y^{-\half})^2 \over y-y^{-1} }(y^{2j_L+1} - y^{-2j_L-1}).}
The value of this index traced over all the quantum states of $N$
black holes gives a measure of the weighted number of supersymmetric
states.

In the supersymmetric quantum mechanics, $j_L$ and $j_R$ are
the eigenvalues of the operators $J^3_L$ (equation \gtn ) and
$J^3_R$ (equation \rch\ specialized to the black hole case),\foot{We recall that
$J_R$ is part of the superconformal algebra, while $J_L$ 
commutes with it.}
augmented by the corresponding operators for the center-of-mass multiplet.
The trace over this latter multiplet gives a universal factor 
of $(y^\half+y^{-\half})^2$.  The total  index
${\cal I}_{BH}$
counting the weighted number of supersymmetric black hole bound states is
then defined by 
\eqn\damk{{\cal I}^{tot}_{BH}=(y^\half+y^{-\half})^2 Tr_{SCQM} (-)^{2J^3_R}y^{2J^3_L},}
where the trace is over the internal
Hilbert space of the
superconformal quantum mechanics, without the center-of-mass multiplet.
We may also define a reduced index of the form \tdf\ by factoring out the 
center of mass factor and restricting to the subspace  
transforming in the dimension $2j_L +1$ representation of $SU(2)_L$,
as
\eqn\radf{{\cal I}_{BH}(j_L)=Tr_{SCQM} (-)^{2J^3_R}|_{j_L}.}
In the next section we will evaluate this for 
two black holes and $j_L=0$ by counting chiral primaries.  

\subsec{$N=2$}

For the case of two black holes, the $D(2,1;0)$ quantum mechanics
governing their relative separation $\vec  X_{12}\equiv \vec
X^1-\vec X^2$ is described by the metric \eqn\nmel{ds^2={12\pi^2(Q_1^2Q_2 +
Q_1Q_2^2)}{d\vec
X_{12} \cdot  d\vec X_{12}\over |\vec X_{12}|^4},}as follows from
\met\ and \gtf. One finds \eqn\dft{ \vec D= -\vec
X_{12},~~~~~~
K={6 \pi^2(Q_1^2 Q_2+Q_2^2 Q_1) \over |\vec X_{12}|^2}, ~~~~~
e^{-\phi}={|\vec
X_{12}|^4 \over 6 \pi^2(Q_1^2 Q_2+Q_2^2 Q_1)}   .}
There are states constructed as in \sdfw\
corresponding to $(p,0)$-forms with $p=0,1,2$. For $p=0$ the
chiral primary condition \chprc\ reduces to $\p f_0=0$ and $D f_0=0$,
which has no nontrivial solutions.

For $p=1$ the condition $D\wedge f_1=0$ implies that $f_1$ is
proportional to $D$. The condition $\p f_1=0$ then implies that the
proportionality factor is a function of $K$ times an 
antiholomorphic function. In complex coordinates $(z^1,z^2)$
\eqn\rthj{f_1=c(\bar z^1, \bar z^2,K)D.}
For the case considered here of $j_L=0$, $f_1$, and therefore 
$c$, must be invariant under $SU(2)_L$ rotations. 
This requires $c$ to be a function of $K$ only. 
Using the formula \bgo\
for the special case $n=2$ and $h=-2$, one finds \eqn\dft{\bar \p
*e^{-\phi}D=0.} It follows that the last condition in \chprc\ is satisfied only when the
proportionality factor is a constant. Hence
the state \eqn\bdfg{|D\rangle=D_a\l^a |0 \rangle} is the unique chiral primary
with $p=1$ for $N=2$. The norm is \eqn\redf{\langle D|D\rangle=2 \int d^{4}X
\sqrt g e^{-\phi}Ke^{-2K}.} For a pair of black holes $K$ goes
like $1 \over r^2$, and so \redf\ converges at both large and small
$r$.\foot{It is now seen explicitly
that the norm $\langle0|0\rangle$ of the $L_0$ ground state
diverges logarithmically at large $r$.
This norm is given by the integral in \redf\ without the factor
of $K$.  }   The state $|D\rangle$ is an $SU(2)_R$ singlet and the $J^3_R=+\half$ element
of an $SU(2)_{\rm Right}$ doublet.

At $p=2$, the first two chiral primary conditions are trivially
satisfied. The general solution of the last condition (using the relations \daz)
is the $(2,0)$-form \eqn\trdc{f(X)I^-_{ab},} where $f(X)$ is holomorphic. \trdc\ is singular unless $f$ is constant.
In this case the norm becomes \eqn\fto{\langle I^-|I^-\rangle=2 \int
d^4X\sqrt{g}e^{-\phi-2K},} where we have used $I^{+ab}I^-_{ab}=2$.
The integral in \fto\ diverges logarithmically at $x \to \infty$. Hence
there are no normalizable chiral primaries at $p=2$.

We conclude that the supersymmetric index ${\cal I}_{BH}(j_L=0)$ is $-1$ for
a pair of black holes.

\centerline {\bf Acknowledgements}
 We are grateful
to J. Maldacena, A. Maloney, J. Michelson and M. Spradlin for
useful conversations, and to M. Stern for pointing out an error in 
an earlier version of this work.
This work is supported in part by DOE grant DE-FG02-91ER40654 and an NDSEG
graduate fellowship.
A. V. is also supported by INTAS-OPEN-97-1312.

\listrefs
\end